\documentclass[a4paper,fleqn,usenatbib]{mnras}

\usepackage{txfonts}
\usepackage{lineno}

\usepackage[T1]{fontenc}
\usepackage{ae,aecompl}

\usepackage[toc,page]{appendix}
\usepackage{natbib}
\usepackage{amssymb}
\usepackage{epsf}
\usepackage{pstricks}
\usepackage{color}
\usepackage{graphicx}
\usepackage{slashed}
\usepackage{relsize}
\usepackage{morefloats}
\usepackage{multirow}
\usepackage{array}
\usepackage{hyperref}
\usepackage[justification=centering]{caption}
\usepackage{url}
\usepackage{upgreek} 

\newcommand{\wcen}{$\omega\ \rm{Cen}$}

\title[Timing of Pulsars in the Globular Cluster Omega Centauri]{Timing of Pulsars in the Globular Cluster Omega Centauri}

\author[S. Dai et al.]
{S. Dai$^{1,2}$\thanks{E-mail: Shi.Dai@westernsydney.edu.au}, S. Johnston$^{2}$, M. Kerr$^{3}$, J. Berteaud$^{4}$, B. Bhattacharyya$^{5}$,
\newauthor F. Camilo$^{6}$, E. Keane$^{7}$\\
$^{1}$School of Science, Western Sydney University, Locked Bag 1797, Penrith South DC, NSW 2751, Australia\\
$^{2}$Australia Telescope National Facility, CSIRO, Space and Astronomy, PO Box 76, Epping, NSW 1710, Australia\\
$^{3}$Space Science Division, Naval Research Laboratory, Washington, DC 20375-5352, USA\\
$^{4}$LAPTh, CNRS, USMB, F-74940 Annecy, France and Univ. Grenoble Alpes, CNRS, IPAG, F-38000 Grenoble, France\\
$^{5}$National Centre for Radio Astrophysics, Tata Institute of Fundamental Research, Pune 411 007, India\\
$^{6}$South African Radio Astronomy Observatory, 2 Fir Street, Observatory 7925, South Africa\\
$^{7}$School of Physics, Trinity College Dublin, College Green, Dublin 2, D02 PN40, Ireland\\
}

\date{Accepted XXX. Received YYY; in original form ZZZ}

\pubyear{2022}

\begin{document}
\label{firstpage}
\pagerange{\pageref{firstpage}--\pageref{lastpage}}
\maketitle

\begin{abstract}
We present the timing of the first five millisecond pulsars discovered in the globular cluster Omega Centauri and the discovery of a pulsar with a spin period of 3.68\,ms. With a timing baseline of $\sim$3.5\,yr we are able to measure the derivative of the spin frequency ($\dot{\nu}$) for the first five pulsars. Upper limits on the pulsar line-of-sight acceleration are estimated and compared with predictions based on analytical models of the cluster. We find that PSRs~J1326$-$4728B and D show large negative accelerations, which are in tension with the minimum acceleration predicted by analytical models. We searched for pulsed $\gamma$-ray signals using 14.3\,yr of data from the \textit{Fermi} Large Area Telescope. Although we found no evidence for $\gamma$-ray pulsations, PSRs~J1326$-$4728A, B, C and E are associated with X-ray sources. This suggests that the observed $\gamma$-ray emission from Omega Centauri is likely caused by the emission of the ensemble of MSPs. Finally, the linearly polarised emission from PSR~J1326$-$4728A yields a rotation measure of $-18\pm8$\,rad\,m$^{-2}$.

\end{abstract}




\section{Introduction} \label{sec:intro}

Long-term timing of radio pulsars in globular clusters (GCs) is a powerful tool to study properties of the core region of GCs which complements other techniques and observations at different wavelengths.  This has been demonstrated for 47Tuc~\citep[e.g.,][]{fcl+01,fck+03,rft+16,frk+17}, Terzan 5~\citep{prf+17} and NGC 6624~\citep{psl+17}, where distances, proper motions and core densities of the GCs were obtained. Limits on the mass of possible black holes in the centre of these GCs have also been obtained using measurements of pulsar accelerations~\citep{frk+17,prf+17,psl+17,apr+18}, in combination with dynamical $N$-body simulations \citep{kbl17,bhs+19}.

Precise measurements of spin, binary and astrometric parameters of pulsars in GCs are also important for the understanding of the formation and evolution of compact objects and the dynamics in the core of GCs. It has been suggested that the number of millisecond pulsars (MSPs) in GCs is correlated with the stellar encounter rate~\citep[e.g.,][]{hct10,bhs+13}. Observed pulsar populations vary amongst GCs, with some primarily comprising binary systems, while others contain more isolated and `exotic' pulsars. \citet{vf14} proposed that these differences are best explained not by the aggregate encounter rate but by the single-object encounter rate, i.e. the rate at which a specific binary system encounters other objects. This rate scales less strongly with the core density. 
More recently, large-scale $N$-body simulations have been used to model the formation and evolution of compact objects in GCs and compared with observed pulsar populations~\citep[e.g.,][]{ykc+19,kyr+20,ykr+22}.  The detection of a fast radio burst from a globular cluster in M81 \citep{Kirsten22} provides further observational evidence for the formation of unusual objects via many-body interactions.

A population of pulsars in the core of Omega Centauri (\wcen\ or NGC~5139), the most massive and luminous GC in the Galaxy, was first discovered by \citet{djk+20}. Five MSPs were detected using the Ultra-Wideband Low receiver~\citep[UWL,][]{hmd+20} of the 64-meter Parkes radio telescope (also known as Murriyang), including one pulsar in an eclipsing binary system with an orbital period of 2.1 hours. Recently, 13 new pulsars were discovered by the MeerKAT radio telescope, which increased the total number of known pulsars in this cluster to 18~\citep{cfr+23}. This opens up opportunities to explore the dynamics of the cluster through radio observations of these pulsars and test the hypothesis of an intermediate-mass black hole in the cluster centre~\citep[e.g.,][]{bhs+19}. Long-term timing of these pulsars allows us to fold $>$10\,yr of $\gamma$-ray photons collected by the Fermi satellite, and thus probe the origin of $\gamma$-ray emission from the core of \wcen~\citep[4FGL~J1326.6$-$4729,][]{aaa+10} and constrain dark matter annihilation models~\citep{bml+19,rbg+21,wby21,sbb+22,cl22}.

In this paper, we present the results of our timing campaign of the first five MSPs discovered in \wcen\ and report the discovery of a further MSP. In $\S$Section~\ref{sec:observations} we describe the timing campaign and observational setup. In $\S$Section~\ref{sec:result} we present the results and discuss their implications in $\S$Section~\ref{sec:discussion}.

\section{Observations and Data Reduction}
\label{sec:observations}

A regular monitoring program of \wcen\ using the Parkes telescope started in 2019, and \wcen\ is currently being observed monthly as part of project P1041 (PI: S. Dai). In our timing analysis below we included observations up to August 2022. Observations were performed using a coherently de-dispersed search mode where data are recorded with 2-bit sampling every 64\,$\upmu$s in each of the 1-MHz wide frequency channels covering a total bandwidth of 3328~MHz between 704 and 4032~MHz. Data were coherently de-dispersed at a DM of 100.3\,${\rm pc\,cm}^{-3}$. Full Stokes information was recorded and a pulsed noise signal with well-known properties injected into the signal path was observed before the observation for the purpose of calibration. For the known pulsars, the search-mode data were folded at the appropriate topocentric spin-period using the {\sc dspsr} software package~\citep{vb11} with a sub-integration length of 30 seconds. We manually excised data affected by narrow-band and impulsive radio-frequency interference for each sub-integration using {\sc psrchive}~\citep{hvm04}. Polarisation and flux calibration were carried out for these search mode observations following \citet{dlb+19}. The linear polarisation and the position angle (PA) of linear polarisation were calculated following \citet{dhm+15}.
Since only PSR J1326$-$4728B can be detected at above 2\,GHz in our individual observations, we only folded search-mode data from 704\,MHz to 2112\,MHz, which corresponds to the lowest 11 subbands of UWL. Each folded file was then averaged in time to form an averaged pulse profile and the pulse time of arrivals (ToAs) was measured for each observation for each pulsar using the \texttt{pat} routine of {\sc psrchive}. Timing analysis was carried out using the {\sc TEMPO2} software package~\citep{hem06}. We fixed the parallax of \wcen\ at $0.191$\,mas~\citep{scr21} throughout our analysis and left spin, astrometric and binary parameters as free parameters.

Periodicity searches were carried out with the software package {\sc presto}~\citep{ran01}. The DM range that we searched was $90-110\,{\rm pc\,cm}^{-3}$. In order to account for possible orbital modulation of pulsar periodic signals, for each 1\,hr segment, we searched for signals drifting by as much as $\pm200/n_{h}$ bins in the Fourier domain by setting the \texttt{zmax} parameter of the \texttt{accelsearch} routine to be $200$~\citep{rem02}, where $n_{h}$ is the largest harmonic at which a signal is detected (up to 8 harmonics were summed). 
%


\section{Results}
\label{sec:result}

\subsection{Discovery of a new MSP}

\begin{figure}
\centering
\includegraphics[width=8cm]{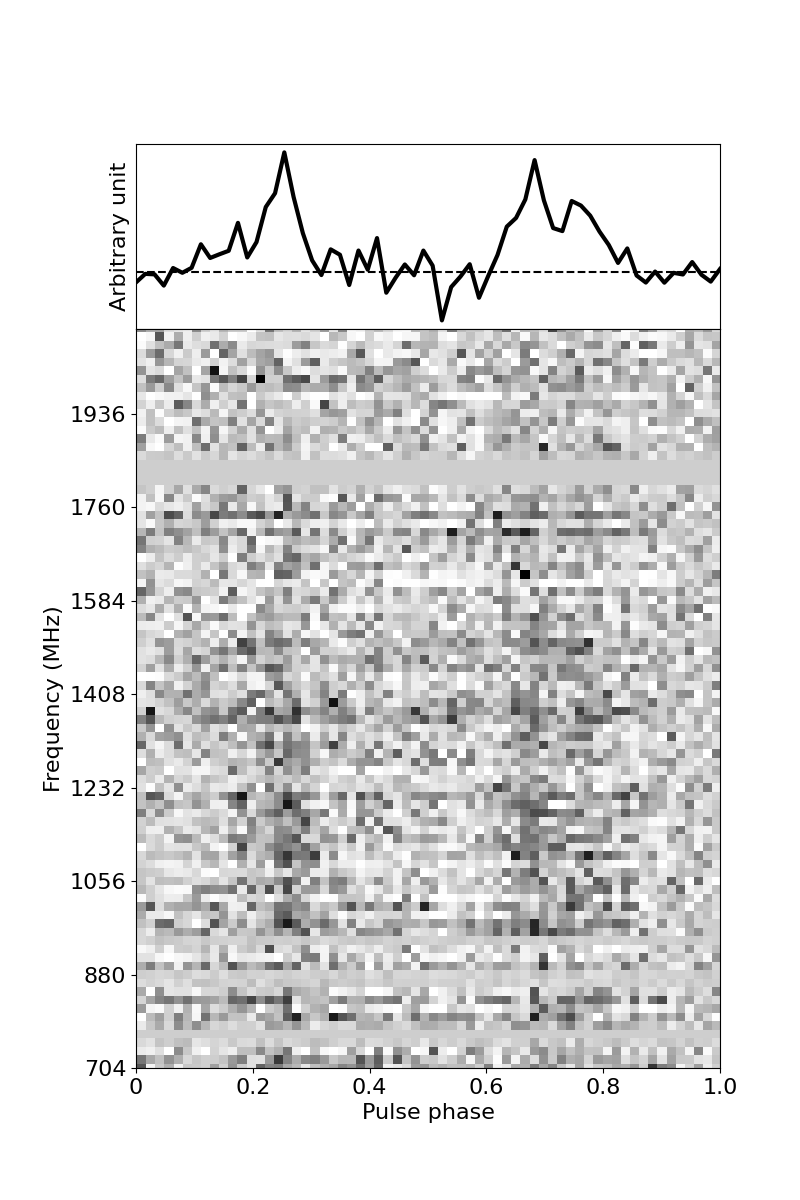}
\caption{Top panel: pulse profile of PSR~J1326$-$4728J averaged across frequencies from 704\,MHz to 2112\,MHz. Bottom panel: frequency spectrum of PSR~J1326$-$4728J. The pulsar cannot be detected above $\sim2$\,GHz.}
\label{fig:psrF}
\end{figure}

We discovered and confirmed a further millisecond pulsar at a DM of 97.3\,${\rm pc\,cm}^{-3}$ with a spin period of 3.686\,ms. In Fig.~\ref{fig:psrF} we show the averaged pulse profile (top) and frequency spectrum (bottom) of the pulsar with an integration time of 3\,hr. This pulsar shows a relatively wide profile with an inter-pulse and can only be detected below 2\,GHz. Limited by the signal-to-noise ratio (S/N) we can achieve with our regular 2\,hr observations, a coherent timing solution of this pulsar is yet to be obtained. Recently, \citet{cfr+23} published the discovery of 13 new pulsars with the MeerKAT radio telescope. One of their discoveries (J1326$-$4728J) has a spin period of 1.84\,ms at a DM of 97.28\,${\rm pc\,cm}^{-3}$. While the S/N of the averaged pulse profile in Fig.~\ref{fig:psrF} is not high, we can see differences in the width of the peaks, and therefore MeerKAT might have discovered J1326$-$4728J at its second harmonic. We suggest that further observations are required to determine the nature of this pulsar.

\subsection{Coherent timing solutions for five pulsars}

\begin{figure}
\centering
\includegraphics[width=8cm]{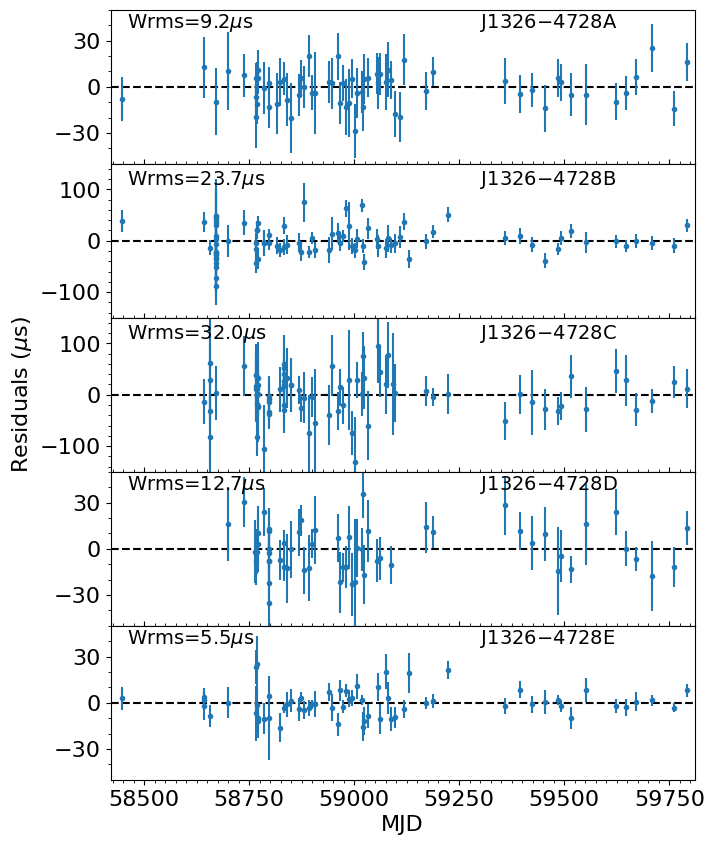}
\caption{Timing residuals of five MSPs with coherent timing solutions.}
\label{fig:resi}
\end{figure}

We obtained coherent timing solutions for all five pulsars published in \citet{djk+20}. Spin, astrometric and binary parameters of these pulsars are listed in Table~\ref{tab:psr}. With a timing baseline of $\sim$3.5\,yr, we measured the first spin frequency derivative ($\dot{\nu}$) and proper motions for all five pulsars, although uncertainties of proper motions are large due to the limited timing baseline. In Fig.~\ref{fig:resi} we show the timing residuals for these pulsars. We achieved timing precision better than $\sim$10\,$\upmu$s (weighted root-mean-square) for PSRs J1326$-$4728A and E.
%

\begin{figure*}
\centering
\includegraphics[width=14cm]{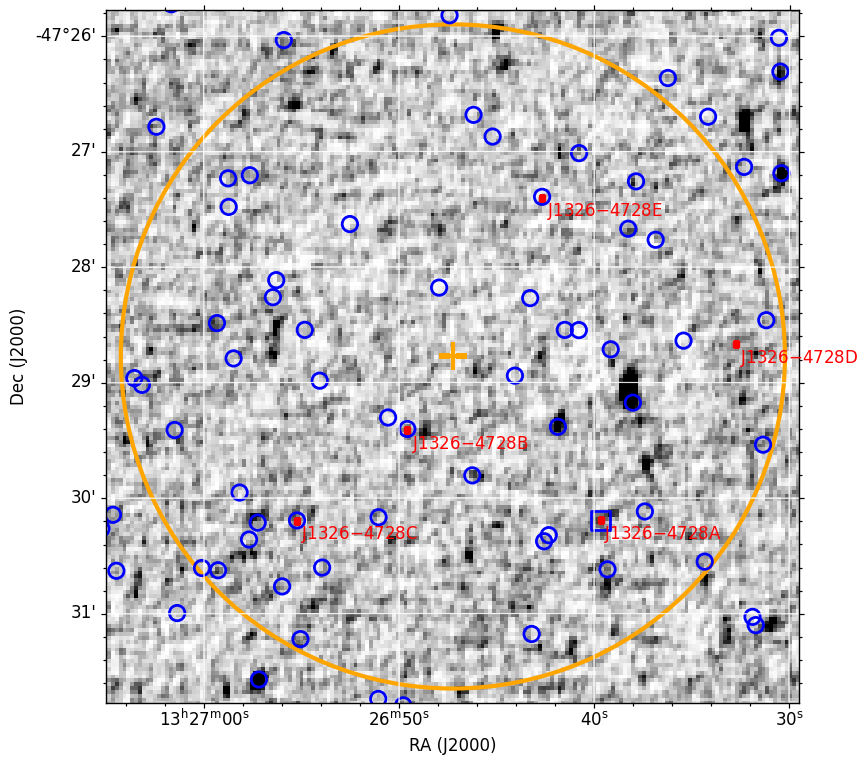}
\caption{Timing positions of \wcen\ pulsars (red squares). ATCA image at 2.1GHz~\citep{djk+20} is shown in the background. 
Blue circles: \textit{Chandra} unidentified sources~\citep{hch+18} with a radius of 4\,arcsec;
Blue square: X-ray counterpart of PSR J1326$-$4728A identified by \citet{zh22}; Orange circle: the core region of \wcen\ centred at RA=13:26:47.24, Dec=$-$47:28:46.5 with a radius of 4.54\,pc~\citep{bh18}.}
\label{fig:atca}
\end{figure*}

In Fig.~\ref{fig:atca} we show the timing positions of these five pulsars superposed on a 2.1-GHz radio continuum obtained with the Australia Telescope Compact Array~\citep{djk+20}. All five pulsars are within one core radius (155\,arcsec) of \wcen. We also show unidentified X-ray sources~\citep{hch+18} in Fig.~\ref{fig:atca} as blue circles. Four pulsars, PSRs~J1326$-$4728A B, C, and E are associated with X-ray sources. X-ray emission from PSR~J1326$-$4728A was detected by \citet{zh22} at its timing position (shown as a blue square in Fig.~\ref{fig:atca}), while the other three pulsars are associated with unidentified X-ray sources published by \citet{hch+18}. The properties of these X-ray sources are listed in Table~\ref{tab:psr}. Compared with detected X-ray emission from MSPs in other GCs~\citep[e.g.,][]{bhc+17}, these X-ray sources show much lower X-ray flux.  

\subsection{Polarisation, DM and spectral index}

\begin{figure*}
\centering
\includegraphics[width=8cm]{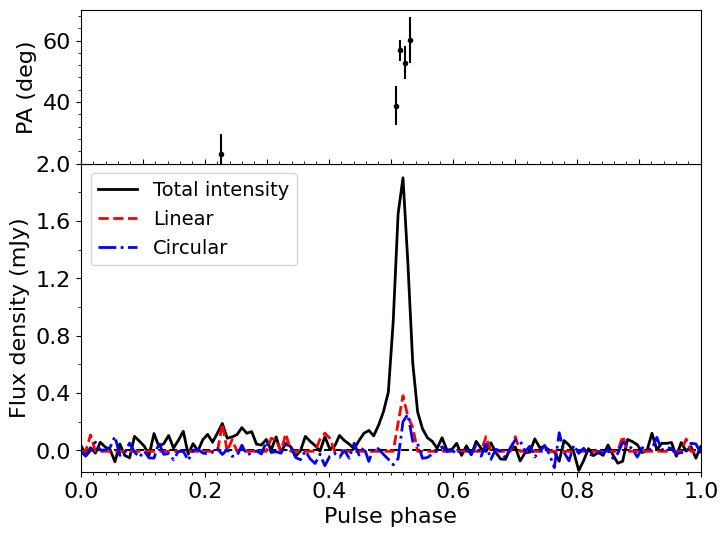}
\includegraphics[width=8cm]{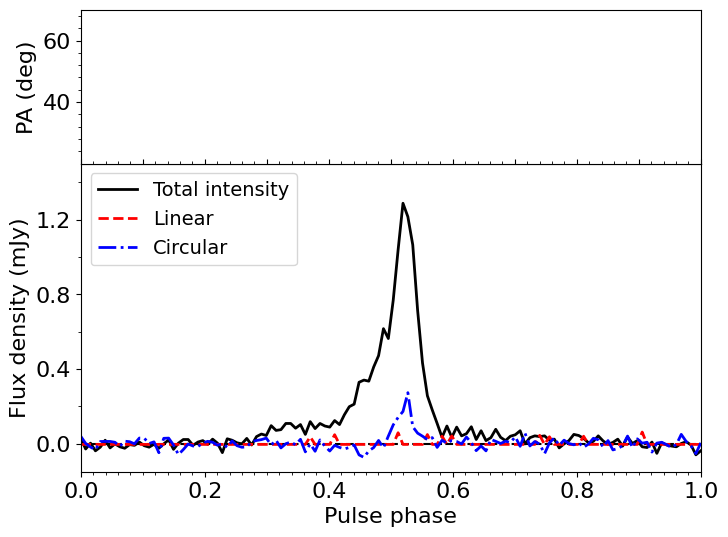}
\caption{Polarisation pulse profiles of J1326$-$4728A (left) and B (right). In bottom panels, the total intensity is shown as black solid lines. Linear polarisation is shown as red dashed lines. Circular polarisation is shown as blue dot-dashed lines. Linear polarisation can be detected only in J1326$-$4728A. In the top panel on the left, we show linear polarisation position angles (PAs) of J1326$-$4728A. }
\label{fig:prof}
\end{figure*}

No linear or circular polarisation was detected in any of these pulsars at their discovery. With three years of observation, we summed all observations of each pulsar using their timing models and formed higher S/N Stokes-parameter profiles. Weak linearly polarised emission was detected in PSR~J1326$-$4728A and we measured a Faraday rotation measure (RM) of $-18\pm8$\,rad\,m$^{-2}$. In Fig.~\ref{fig:prof}, we show the polarisation profiles of PSR~J1326$-$4728A and B, in which circularly polarised emission can be observed.

We averaged these high S/N profiles into eight frequency channels (with a channel bandwidth of 176\,MHz) and then measured the flux density ($S$) and ToA for each channel. We used {\sc TEMPO2} to fit the DM for each pulsar and also fitted the spectrum of each pulsar with a single power law ($S=S_{0}\nu^{\alpha}$, where $\alpha$ is the spectral index). The refined DM measurement, spectral index, and flux density (at a central frequency of 1489\,MHz) of each pulsar are listed in Table~\ref{tab:psr}. 

\subsection{$\gamma$-ray emission}

To search for pulsed $\gamma$-ray signals, we used Pass 8 \citep{Pass8,improvedPass8} data from the \textit{Fermi} Large Area Telescope \citep{LATinstrument} selected with energies 0.1--30\,GeV within 3$^{\circ}$ of the cluster center and with a time range MJD 54682--59900, or 14.3\,yr.  With version 2.0.0 of the fermitools\footnote{\url{https://fermi.gsfc.nasa.gov/ssc/data/analysis/software/}}, we computed photon weights \citep{KerrWeighted} using the 4FGL-DR3 \citep{4FGL-DR3} model.  We then assigned spin phase to (``folded'') each photon using the five timing solutions and the \textit{Fermi} plugin \citep{Ray2011} for \textsc{Tempo2}.  We calculated the H-test \citep{Htest} statistic both with and without photon weights and in time ranges restricted to the ephemeris validity and for the full dataset.  The largest observed value was 12, which has a chance probability of 0.01.  Considering the number of combinations and pulsars tested, this outcome is consistent with a uniform distribution of pulse phases, i.e. there is no evidence for pulsations.

\begin{table*}
\begin{center}
\caption{Parameters of five pulsars in \wcen. Flux densities were measured at a central frequency of 1489\,MHz ($S_{1489}$). X-ray source IDs are from \citet{hch+18}. For J1326$-$4728A, X-ray emission was extracted from a circular region with a 1\,arcsec radius centred on the radio timing positions~\citep{zh22}.  The listed companion masses for J1326$-$4728B are the minimum, median, and maximum values.}
\label{tab:psr}
\begin{tabular}{lccccc}
\hline
\hline
          & J1326$-$4728A          &  J1326$-$4728B        & J1326$-$4728C   &  J1326$-$4728D  & J1326$-$4728E\\
\hline
RAJ (J2000)        & 13:26:39.6699(2)         &  13:26:49.5688(3)      & 13:26:55.2219(6)   & 13:26:32.7130(2)   & 13:26:42.67844(7)    \\
DECJ (J2000)       & $-$47:30:11.641(3)       &  $-$47:29:24.889(4)    & $-$47:30:11.753(9) & $-$47:28:40.053(3) & $-$47:27:23.999(1)  \\
$\nu$ (Hz)         & 243.38088019784(7)       & 208.68683313306(7)     & 145.6057701815(1)  & 218.39623714185(2) & 237.65856646672(2)  \\
$\dot{\nu}$ (Hz/s) & $-1.622(1)\times10^{-15}$& $2.366(2)\times10^{-15}$   & $-2.08(2)\times10^{-16}$  & $1.960(2)\times10^{-15}$ & $-9.194(6)\times10^{-16}$ \\
PMRA (mas/yr)      & -5(1) & -3(3)   &  -5(3)  & -3(2)  &  -4.2(7)                        \\
PMDEC (mas/yr)     & -7(2) & -8(4)   &  -4(5)  & -6(3)  &  -6.9(9)                        \\
PEPOCH (MJD)       &  58447.77             &  58768.0              &     58447.77           & 58797.01            &  58796.79 \\
Time span (MJD)    &  58447.80$-$59792.18  &  58447.78$-$59792.18  &   58643.33$-$59792.18  & 58700.27$-$59792.18 &  58447.79$-$59792.18 \\
DM (cm$^{-3}$\,pc)  &  100.2899(7)         &  100.250(1)      &    100.640(5)   & 96.518(2)     &  94.3690(4) \\
RM (rad\,m$^{-2}$) & -18(8) &  &  &  & \\
$S_{1489}$ (mJy) & 54(5) & 88(4) & 31(3) & 38(4) & 33(2) \\
\hline
\multicolumn{6}{c}{Binary parameters~\citep[ELL1 model,][]{hem06}}\\
\hline
$P_{\rm{b}}$ (days) &         &  0.0896112039(9)     &       &         \\
$\chi$ (ls)         &         &  0.021450(4)         &       &  \\
$T_{\rm{asc}}$ (MJD) &        &  58768.037248(4)     &       &\\
$\eta$ (10$^{-3}$)   &        &  -0.6(4)             &       &  \\
$\kappa$ (10$^{-3}$) &        &  -0.6(4)             &       & \\
\hline
\multicolumn{6}{c}{Derived parameters}\\
\hline
$P$ (ms)                 & 4.108786192190(1)        &  4.791869161014(2)          &  6.867859692327(6)        & 4.578833468410(2)         & 4.2077170407405(4) \\
$\dot{P}$ (s\,s$^{-1}$)  & $2.738(2)\times10^{-20}$ &  $-5.433(4)\times10^{-20}$  &  $9.8(1)\times10^{-21}$   & $-4.110(4)\times10^{-20}$ & $1.628(1)\times10^{-20}$\\
Spectral index     & $-1.9(1)$ & $-0.7(1)$ & $-1.9(2)$ & $-2.3(2)$  & $-2.3(3)$ \\
Companion mass (M$_{\odot}$)  & & 0.013 < 0.016 < 0.031 & & \\
\hline
\multicolumn{6}{c}{Association with X-ray sources~\citep{hch+18,zh22}}\\
\hline
X-ray source ID&               &   13d            & 23g            &  &  11f        \\
RAJ (J2000)    & 13:26:39.670  &  13:26:49.574    & 13:26:55.231   &  & 13:26:42.670\\
DECJ (J2000)   & $-$47:30:11.64 &  $-$47:29:24.18  & $-$47:30:11.63 &  & $-$47:27:23.56\\
Flux ($10^{-16}$\,erg\,cm$^{-2}$\,s$^{-1}$) & 6.1 & 20.3 & 9.2 & & 6.3 \\
\hline
\end{tabular}
\end{center}
\end{table*}


\section{Discussion and conclusion}
\label{sec:discussion}

The observed spin-down ($\dot{P}_{\rm obs}$) of pulsars in GCs is the sum of several effects and is typically dominated by the dynamical effect caused by the cluster gravitational potential. Here we briefly discuss the contribution from these various effects for pulsars in \wcen. More details on these effects can be found in previous papers~\citep[e.g.,][]{frk+17,prf+17,psl+17,apr+18}.  
For pulsars located in the same GC, their $\dot{P}_{\rm obs}$ are affected by two types of common line-of-sight accelerations: 
\begin{itemize}
    \item the difference between the line-of-sight accelerations of the Solar system and the GC in the field of the Galaxy ($a_{\rm g}$). Following Section 5.1.2 of \citet{prf+17}, we estimated this contribution to be
    \begin{equation}
        \label{accl_g}
        a_{\rm g}=-1.36\times10^{-10}\,\rm{m}\,\rm{s}^{-2}
    \end{equation} in the direction of \wcen\ and at a distance of 5.43\,kpc~\citep{bv21}. 
    \item the apparent acceleration due to the proper motion of pulsars ($a_{\rm s}$), or the so-called Shklovskii effect~\citep{shk70}. The Shklovskii effect can be estimated as  
    \begin{equation}
    \label{accl_s}
        a_{\rm s}=\mu^{2}_{\rm T}d=3.95\times10^{-12}\left(\frac{d}{5.43\,\rm{kpc}}\right)\left(\frac{\mu_{\rm T}}{\rm{mas}\,\rm{yr}^{-1}}\right)^2\,\rm{m}\,\rm{s}^{-2},
    \end{equation}
    where $d$ is the distance to the cluster, $\mu_{\rm T}=\sqrt{\mu_{\delta}^{2}+\mu_{\alpha *}^{2}}$ is the proper motion. Since uncertainties of pulsar proper motions are large due to the limited timing baseline, we used the mean proper motion of \wcen, $\mu_{\alpha *}=-3.1925$\,mas\,yr$^{-1}$,  $\mu_{\delta}=-6.7445$\,mas\,yr$^{-1}$~\citep{hvm+18}, and estimated the acceleration due to Shklovskii effect to be $a_{\rm s}\approx2.2\times10^{-10}$\,m\,s$^{-2}$. The contribution to the Shklovskii effect from pulsar velocities relative to the GC should be negligible since the escape velocity in the core of \wcen\ is measured to be only $\sim60$\,km\,s$^{-1}$~\citep{gzp+02}. At a distance of 5.43\,kpc, this corresponds to a maximal projected proper motion of $\sim2.3$\,mas\,yr$^{-1}$ with respect to the cluster centre, which is significantly smaller than the mean proper motion of \wcen. 
\end{itemize}

In addition to those common effects, there are also pulsar-dependent contributions to $\dot{P}_{\rm obs}$, including: 
\begin{itemize}
    \item the line-of-sight acceleration of the pulsar in the gravitational field of the GC ($a_{l,\rm{GC}}$), which depends on the location of the pulsar with respect to the centre of the GC and is typically unknown.
    \item the intrinsic pulsar spin-down ($\dot{P}_{\rm int}$). For pulsars in GCs, $\dot{P}_{\rm int}$ can not be directly measured due to the dynamical effects caused by the gravitational potential and is therefore unknown. However, an order of magnitude estimate can be made by assuming a simple magnetic dipole emission model with a braking index of $n=3$. With these assumptions, the apparent acceleration due to intrinsic spin-down $\dot{P}_{\rm int}$ can be estimated as
    \begin{equation}
        a_{\rm int}=c\left(\frac{\dot{P}}{p}\right)=7.96\times10^{-10}\left(\frac{B}{2\times10^{8}\,{\rm G}}\right)^{2}\left(\frac{2\,{\rm ms}}{P}\right)^{2}\,\rm{m\,s^{-2}},
    \end{equation}
    where $B$ is the surface magnetic field strength of the pulsar. 
    \item the acceleration caused by individual objects near the pulsar, which may be appreciable due to the high stellar density in the core region of GCs. While it is difficult to estimate this acceleration for a given pulsar, the ratio of the characteristic nearest-neighbour acceleration, $a_{\rm NN}$, to the mean cluster acceleration for a cluster of mass $M_{\rm cl}\sim Nm$ and radius $R_{\rm cl}\sim (N/n)^{1/3}$ and $N$ stars of mass $m$ can be estimated as 
    \begin{equation}
        \frac{a_{\rm NN}}{GM_{\rm cl}/R^{2}_{\rm cl}} \sim 10^{-2}\left(\frac{10^{6}}{N}\right)^{1/3},
    \end{equation}
    where $n$ is the mean number density~\citep{prf+17}. For \wcen, the projected number density is $\sim10^{6}$ stars in the central region, and therefore accelerations caused by nearby stars are negligible compared with the mean cluster acceleration and we will not include this term in the following discussions.
\end{itemize}

\begin{figure}
\centering
\includegraphics[width=8cm]{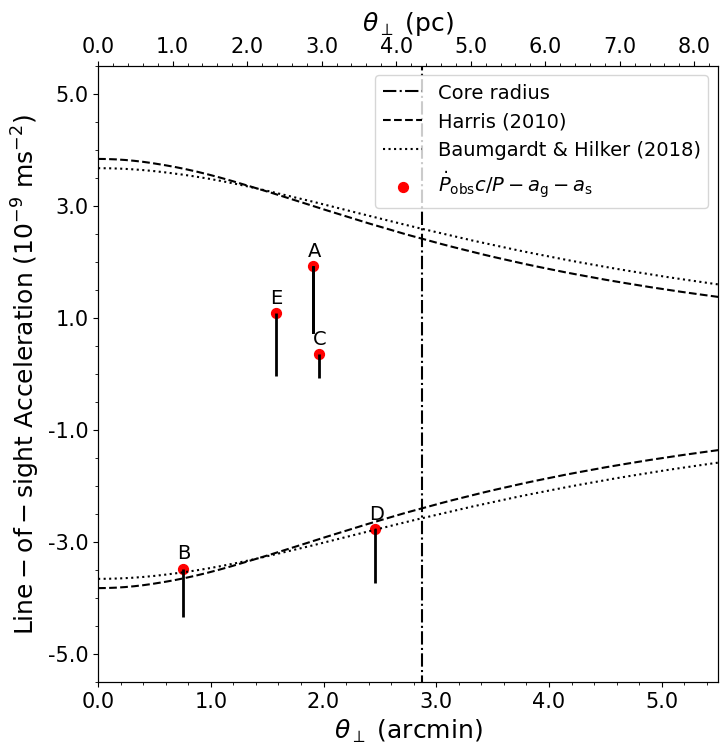}
\caption{Line-of-sight acceleration as a function of the angular offset from the GC centre. The dash-dotted line shows the core radius of $r_{\rm c}=4.54$\,pc~\citep{bh18}. Red points show accelerations after correcting for contributions from the Galactic potential and Shklovskii effect (using the mean proper motion of the cluster). The vertical solid line of each point shows the contribution of intrinsic spin-down assuming a surface magnetic field strength of $5\times10^8$\,G. Dashed and dotted lines show the maximum and minimum accelerations along each line of sight predicted by an analytical model of the cluster using cluster parameters from \citet{har10} and \citet{bh18} (see Section~\ref{sec:discussion} for details). The distance to the cluster is fixed at 5.43\,kpc~\citep{bv21}.}
\label{fig:accel}
\end{figure}

Taking all these effects into account, the observed spin-down of pulsars in GCs can be expressed as 
\begin{equation}
    \dot{P}_{\rm obs} = \frac{(a_{l,\rm{GC}}+\mu^{2}_{\rm T}d+a_{\rm g})P}{c}+\dot{P}_{\rm int}, 
\end{equation}
where $P$ is the observed pulsar spin period and $c$ is the speed of light. Since $a_{l,\rm{GC}}$ and $\dot{P}_{\rm int}$ are unknown for most pulsars in GCs, we can only put an upper limit on the pulsar line-of-sight acceleration as
\begin{equation}
\label{accl}
    a_{l,\rm{max}}\doteq a_{l,\rm{GC}}+\frac{\dot{P}_{\rm int}}{P}c=\frac{\dot{P}_{\rm obs}}{P}c-\mu^{2}_{\rm T}d-a_{\rm g}.
\end{equation}
Upper limits on the pulsar line-of-sight acceleration can then be compared with the maximum and minimum accelerations along each line of sight predicted by GC density models. 
Following \citet{frk+17} and assuming the \citet{kin62} density profile, we calculated the maximum and minimum accelerations as, 
\begin{equation}
    a_{\rm GC}(x)=\frac{9\sigma_{\mu,0}^{2}d}{\theta_{\rm c}}\frac{1}{x^{2}}\left(\frac{x}{\sqrt{1+x^{2}}}-{\rm sinh^{-1}}x\right),
\end{equation}
where $x$ is the distance to the centre divided by the core radius, $\sigma_{\mu,0}$ is the central velocity dispersion and $\theta_{\rm c}$ is the core radius in radian. The distance to the cluster is $d=5.43$\,kpc~\citep{bv21}. 
In Fig.~\ref{fig:accel} we show the line-of-sight acceleration for each pulsar as a function of their angular offset from the GC centre. Red points show accelerations after correcting for contributions from the Galactic potential (Eq.~\ref{accl_g}) and Shklovskii effect (Eq.~\ref{accl_s}) based on the mean proper motion of the cluster. The vertical solid line of each pulsar shows the contribution from the intrinsic spin-down assuming a surface magnetic field strength of $5\times10^8$\,G, a typical value for pulsars in GCs~\citep[e.g.,][]{fck+03,prf+17}. The dashed and dotted lines show the maximum and minimum accelerations calculated using \wcen\ core radius and central velocity dispersion measurements from \citet{har10} and \citet{bh18}. \citet{har10} measured the core radius and central velocity dispersion to be $\theta_{\rm c}=2.37$\arcmin\ and $16.8$\,km/s, respectively. More recently, \citet{bh18} presented new measurements of 112 Galactic GCs by fitting a large set of $N$-body simulations to their velocity dispersion and surface density profiles. They measured the core radius and central velocity dispersion of \wcen\ to be $r_{\rm c}=4.54$\,pc (or 2.88\arcmin) and $18.1$\,km/s, respectively. 


%


Fig.~\ref{fig:accel} shows that the line-of-sight acceleration of PSRs J1326$-$4728A, C and E are within the maximum and minimum acceleration predicted by the analytic model. Their positive accelerations suggest that they are located on the far side of the cluster centre. PSRs J1326$-$4728B and D show large negative accelerations, which are in tension with the minimum acceleration predicted by the analytical model. PSRs J1326$-$4728B is also the closest pulsar to the cluster centre with an angular offset $<$1$\arcmin$ (or $\sim$1.1\,pc). As shown in Fig.~\ref{fig:accel}, even without considering the pulsar intrinsic spin-down, the line-of-sight acceleration of J1326$-$4728B and D are already close to the minimum acceleration predicted by the analytical model. Taking a moderate surface magnetic field of $5\times10^8$\,G into account, the line-of-sight acceleration of PSRs J1326$-$4728B and D significantly exceed minimum accelerations estimated using both \citet{har10} and \citet{bh18}. Recent inferences of the magnetic field configuration of some MSPs \citep{Kalapotharakos21} suggest the presence of asymmetric magnetic fields which can contribute acceleration through the ``rocket effect'' \citep{Harrison75}.  However, such acceleration is $<$$10^{-12}(\dot{E}/10^{34}\mathrm{erg\,s^{-1}})$\,m\,s$^{-2}$, which is again too low to contribute a dynamical effect for reasonable intrinsic spin-down rates.

The single-mass, non-rotating, isotropic \citet{kin62} model is clearly oversimplified for \wcen\ considering its large and extended core. Recent work suggests the existence of a centrally concentrated cluster of stellar-mass black holes in \wcen\ to explain its surface brightness and velocity dispersion profile~\citep[e.g.,][]{zgh19,bhs+19}. Using an axisymmetric dynamical model, \citet{esz22} found evidence for a centrally-concentrated distribution of matter that is distinct from the luminous component, with a mass of 10$^4$ to 10$^6$\,M$_{\odot}$. Whether the observed large line-of-sight accelerations of PSR~J1326$-$4728B and D can be understood requires more realistic models of \wcen. Further studies of these pulsars, for example in combination with $N$-body simulations~\citep{bhs+19}, could potentially provide new insights into the existence and properties of stellar-mass black holes and/or an intermediate-mass black hole in \wcen. Continued timing of these pulsars and precise measurements of their proper motions, higher order spin frequency derivatives and the orbital period derivative of J1326$-$4728B will be crucial for us to probe the dynamics in the core of \wcen.

So far, we have discovered five isolated pulsars and only one binary system in \wcen. Our Parkes survey is likely biased towards isolated pulsars due to limited sensitivity and relatively long integration time ($\sim2.5$\,hr). With the more sensitive MeerKAT radio telescope, \citet{cfr+23} discovered four new pulsars in compact binary systems and another three pulsars in potentially wide binaries. 
Even considering these new discoveries, as discussed by \citet{cfr+23}, the fraction of isolated pulsars in \wcen\ is surprisingly high considering its extended core and therefore low encounter rate for a single binary~\citep{vf14}. Deeper surveys with sensitive telescopes are therefore necessary and could reveal if the pulsar population in \wcen\ has a different evolutionary history or significantly different properties. 


Finally, PSR~1326$-$4728A, B, C and E are shown to be associated with X-ray sources~\citep{hch+18,zh22}. Considering that no $\gamma$-ray pulsation has been detected, this strongly indicates that the observed $\gamma$-ray emission is the summed emission of the ensemble of MSPs. PSR~J1326$-$4728B is the brightest in X-ray and \citet{zh22} showed that its X-ray spectrum can be well fitted by a single power-law, which gives an unabsorbed X-ray flux of $3.3\pm0.9\times10^{-15}$\,erg\,cm$^{-2}$\,s$^{-1}$ and a photon index of $2.6\pm0.5$. Using the $\gamma$-to-X flux relation presented by \citet{bcc+21}, we estimated the $\gamma$-ray emission from PSR~J1326$-$4728B to be $\sim3.2\times10^{-12}$\,erg\,cm$^{-2}$\,s$^{-1}$ (0.1 to 100\,GeV), about 30\% of the observed $\gamma$-ray flux from \wcen~\citep[$1.1\times10^{-11}$\,erg\,cm$^{-2}$\,s$^{-1}$,][]{4FGL-DR3}. A recent analysis by \citet{smh+21} suggested that inverse Compton emission from relativistic pairs launched by MSPs in \wcen\ could make a significant contribution to the observed $\gamma$-ray emission.

\section*{Acknowledgements}

We thank A. Possenti, M. Burgay and V. Morello for helping with observations and data analysis. We thank H. Baumgardt for the useful discussions.
The Parkes radio telescope is part of the Australia Telescope National Facility which is funded by the Commonwealth of Australia for operation as a National Facility managed by CSIRO. S.D. is the recipient of an Australian Research Council Discovery Early Career Award (DE210101738) funded by the Australian Government.
J.B. acknowledges financial support from the Programme National des Hautes Energies of CNRS/INSU with INP and IN2P3, co-funded by CEA and CNES, from the “Agence Nationale de la Recherche”, grant n. ANR-19-CE31-0005-01 (PI: F. Calore), and from the Centre National d’Etudes Spatiales (CNES)
B.B. acknowledges the Department of Atomic Energy, Government of India, under project no.12-R\&D-TFR-5.02-0700
Work at NRL is supported by NASA.

The {\it Fermi}-LAT Collaboration acknowledges generous ongoing support from a number of agencies and institutes that have supported both the development and the operation of the LAT as well as scientific data analysis. These include the National Aeronautics and Space Administration and the Department of Energy in the United States, the Commissariat \`a l'Energie Atomique and the Centre National de la Recherche Scientifique / Institut National de Physique Nucl\'eaire et de Physique des Particules in France, the Agenzia Spaziale Italiana and the Istituto Nazionale di Fisica Nucleare in Italy, the Ministry of Education, Culture, Sports, Science and Technology (MEXT), High Energy Accelerator Research Organization (KEK) and Japan Aerospace Exploration Agency (JAXA) in Japan, and the K.~A.~Wallenberg Foundation, the Swedish Research Council and the Swedish National Space Board in Sweden.\newline

Additional support for science analysis during the operations phase is gratefully acknowledged from the Istituto Nazionale di Astrofisica in Italy and the Centre National d'\'Etudes Spatiales in France. This work was performed in part under DOE Contract DE-AC02-76SF00515. \newline

\section*{Data availability}
The observations from the Parkes radio telescope are publicly available from \url{https://data.csiro.au/} after an 18-month embargo period. Pulse time of arrivals will be shared on reasonable request.

\bibliography{ms}

\begin{thebibliography}{}
\makeatletter
\relax
\def\mn@urlcharsother{\let\do\@makeother \do\$\do\&\do\#\do\^\do\_\do\%\do\~}
\def\mn@doi{\begingroup\mn@urlcharsother \@ifnextchar [ {\mn@doi@}
  {\mn@doi@[]}}
\def\mn@doi@[#1]#2{\def\@tempa{#1}\ifx\@tempa\@empty \href
  {http://dx.doi.org/#2} {doi:#2}\else \href {http://dx.doi.org/#2} {#1}\fi
  \endgroup}
\def\mn@eprint#1#2{\mn@eprint@#1:#2::\@nil}
\def\mn@eprint@arXiv#1{\href {http://arxiv.org/abs/#1} {{\tt arXiv:#1}}}
\def\mn@eprint@dblp#1{\href {http://dblp.uni-trier.de/rec/bibtex/#1.xml}
  {dblp:#1}}
\def\mn@eprint@#1:#2:#3:#4\@nil{\def\@tempa {#1}\def\@tempb {#2}\def\@tempc
  {#3}\ifx \@tempc \@empty \let \@tempc \@tempb \let \@tempb \@tempa \fi \ifx
  \@tempb \@empty \def\@tempb {arXiv}\fi \@ifundefined
  {mn@eprint@\@tempb}{\@tempb:\@tempc}{\expandafter \expandafter \csname
  mn@eprint@\@tempb\endcsname \expandafter{\@tempc}}}

\bibitem[\protect\citeauthoryear{{Abbate}, {Possenti}, {Ridolfi}, {Freire},
  {Camilo}, {Manchester}  \& {D'Amico}}{{Abbate} et~al.}{2018}]{apr+18}
{Abbate} F.,  {Possenti} A.,  {Ridolfi} A.,  {Freire} P.~C.~C.,  {Camilo} F.,
  {Manchester} R.~N.,   {D'Amico} N.,  2018, \mn@doi [\mnras]
  {10.1093/mnras/sty2298}, \href
  {https://ui.adsabs.harvard.edu/abs/2018MNRAS.481..627A} {481, 627}

\bibitem[\protect\citeauthoryear{{Abdo} et~al.,}{{Abdo} et~al.}{2010}]{aaa+10}
{Abdo} A.~A.,  et~al., 2010, \mn@doi [\aap] {10.1051/0004-6361/201014458},
  \href {https://ui.adsabs.harvard.edu/abs/2010A&A...524A..75A} {524, A75}

\bibitem[\protect\citeauthoryear{{Abdollahi} et~al.,}{{Abdollahi}
  et~al.}{2022}]{4FGL-DR3}
{Abdollahi} S.,  et~al., 2022, \mn@doi [\apjs] {10.3847/1538-4365/ac6751},
  \href {https://ui.adsabs.harvard.edu/abs/2022ApJS..260...53A} {260, 53}

\bibitem[\protect\citeauthoryear{{Atwood} et~al.,}{{Atwood}
  et~al.}{2009}]{LATinstrument}
{Atwood} W.~B.,  et~al., 2009, \mn@doi [\apj] {10.1088/0004-637X/697/2/1071},
  \href {http://adsabs.harvard.edu/abs/2009ApJ...697.1071A} {697, 1071}

\bibitem[\protect\citeauthoryear{{Atwood} et~al.,}{{Atwood}
  et~al.}{2013}]{Pass8}
{Atwood} W.,  et~al., 2013, arXiv e-prints, \href
  {https://ui.adsabs.harvard.edu/abs/2013arXiv1303.3514A} {p. arXiv:1303.3514}

\bibitem[\protect\citeauthoryear{{Bahramian}, {Heinke}, {Sivakoff}  \&
  {Gladstone}}{{Bahramian} et~al.}{2013}]{bhs+13}
{Bahramian} A.,  {Heinke} C.~O.,  {Sivakoff} G.~R.,   {Gladstone} J.~C.,  2013,
  \mn@doi [\apj] {10.1088/0004-637X/766/2/136}, \href
  {https://ui.adsabs.harvard.edu/abs/2013ApJ...766..136B} {766, 136}

\bibitem[\protect\citeauthoryear{{Baumgardt} \& {Hilker}}{{Baumgardt} \&
  {Hilker}}{2018}]{bh18}
{Baumgardt} H.,  {Hilker} M.,  2018, \mn@doi [\mnras] {10.1093/mnras/sty1057},
  \href {https://ui.adsabs.harvard.edu/abs/2018MNRAS.478.1520B} {478, 1520}

\bibitem[\protect\citeauthoryear{{Baumgardt} \& {Vasiliev}}{{Baumgardt} \&
  {Vasiliev}}{2021}]{bv21}
{Baumgardt} H.,  {Vasiliev} E.,  2021, \mn@doi [\mnras]
  {10.1093/mnras/stab1474}, \href
  {https://ui.adsabs.harvard.edu/abs/2021MNRAS.505.5957B} {505, 5957}

\bibitem[\protect\citeauthoryear{{Baumgardt} et~al.,}{{Baumgardt}
  et~al.}{2019}]{bhs+19}
{Baumgardt} H.,  et~al., 2019, \mn@doi [\mnras] {10.1093/mnras/stz2060}, \href
  {https://ui.adsabs.harvard.edu/abs/2019MNRAS.488.5340B} {488, 5340}

\bibitem[\protect\citeauthoryear{{Berteaud}, {Calore}, {Clavel}, {Serpico},
  {Dubus}  \& {Petrucci}}{{Berteaud} et~al.}{2021}]{bcc+21}
{Berteaud} J.,  {Calore} F.,  {Clavel} M.,  {Serpico} P.~D.,  {Dubus} G.,
  {Petrucci} P.-O.,  2021, \mn@doi [\prd] {10.1103/PhysRevD.104.043007}, \href
  {https://ui.adsabs.harvard.edu/abs/2021PhRvD.104d3007B} {104, 043007}

\bibitem[\protect\citeauthoryear{{Bhattacharya}, {Heinke}, {Chugunov},
  {Freire}, {Ridolfi}  \& {Bogdanov}}{{Bhattacharya} et~al.}{2017}]{bhc+17}
{Bhattacharya} S.,  {Heinke} C.~O.,  {Chugunov} A.~I.,  {Freire} P. C.~C.,
  {Ridolfi} A.,   {Bogdanov} S.,  2017, \mn@doi [\mnras]
  {10.1093/mnras/stx2241}, \href
  {https://ui.adsabs.harvard.edu/abs/2017MNRAS.472.3706B} {472, 3706}

\bibitem[\protect\citeauthoryear{{Brown}, {Massey}, {Lacroix}, {Strigari},
  {Fattahi}  \& {B{\oe}hm}}{{Brown} et~al.}{2019}]{bml+19}
{Brown} A.~M.,  {Massey} R.,  {Lacroix} T.,  {Strigari} L.~E.,  {Fattahi} A.,
  {B{\oe}hm} C.,  2019, arXiv e-prints, \href
  {https://ui.adsabs.harvard.edu/abs/2019arXiv190708564B} {p. arXiv:1907.08564}

\bibitem[\protect\citeauthoryear{{Bruel}, {Burnett}, {Digel}, {Johannesson},
  {Omodei}  \& {Wood}}{{Bruel} et~al.}{2018}]{improvedPass8}
{Bruel} P.,  {Burnett} T.~H.,  {Digel} S.~W.,  {Johannesson} G.,  {Omodei} N.,
   {Wood} M.,  2018, 8th Internat'l Fermi Symposium, \href
  {https://ui.adsabs.harvard.edu/abs/2018arXiv181011394B} {p. arXiv:1810.11394}

\bibitem[\protect\citeauthoryear{{Chan} \& {Lee}}{{Chan} \& {Lee}}{2022}]{cl22}
{Chan} M.~H.,  {Lee} C.~M.,  2022, \mn@doi [\prd]
  {10.1103/PhysRevD.105.123006}, \href
  {https://ui.adsabs.harvard.edu/abs/2022PhRvD.105l3006C} {105, 123006}

\bibitem[\protect\citeauthoryear{{Chen} et~al.,}{{Chen} et~al.}{2023}]{cfr+23}
{Chen} W.,  et~al., 2023, arXiv e-prints, \href
  {https://ui.adsabs.harvard.edu/abs/2023arXiv230103864C} {p. arXiv:2301.03864}

\bibitem[\protect\citeauthoryear{{Dai} et~al.,}{{Dai} et~al.}{2015}]{dhm+15}
{Dai} S.,  et~al., 2015, \mn@doi [\mnras] {10.1093/mnras/stv508}, \href
  {https://ui.adsabs.harvard.edu/abs/2015MNRAS.449.3223D} {449, 3223}

\bibitem[\protect\citeauthoryear{{Dai} et~al.,}{{Dai} et~al.}{2019}]{dlb+19}
{Dai} S.,  et~al., 2019, \mn@doi [\apjl] {10.3847/2041-8213/ab0e7a}, \href
  {https://ui.adsabs.harvard.edu/abs/2019ApJ...874L..14D} {874, L14}

\bibitem[\protect\citeauthoryear{{Dai}, {Johnston}, {Kerr}, {Camilo},
  {Cameron}, {Toomey}  \& {Kumamoto}}{{Dai} et~al.}{2020}]{djk+20}
{Dai} S.,  {Johnston} S.,  {Kerr} M.,  {Camilo} F.,  {Cameron} A.,  {Toomey}
  L.,   {Kumamoto} H.,  2020, \mn@doi [\apjl] {10.3847/2041-8213/ab621a}, \href
  {https://ui.adsabs.harvard.edu/abs/2020ApJ...888L..18D} {888, L18}

\bibitem[\protect\citeauthoryear{{Evans}, {Strigari}  \& {Zivick}}{{Evans}
  et~al.}{2022}]{esz22}
{Evans} A.~J.,  {Strigari} L.~E.,   {Zivick} P.,  2022, \mn@doi [\mnras]
  {10.1093/mnras/stac261}, \href
  {https://ui.adsabs.harvard.edu/abs/2022MNRAS.511.4251E} {511, 4251}

\bibitem[\protect\citeauthoryear{{Freire}, {Camilo}, {Lorimer}, {Lyne},
  {Manchester}  \& {D'Amico}}{{Freire} et~al.}{2001}]{fcl+01}
{Freire} P.~C.,  {Camilo} F.,  {Lorimer} D.~R.,  {Lyne} A.~G.,  {Manchester}
  R.~N.,   {D'Amico} N.,  2001, \mn@doi [\mnras]
  {10.1046/j.1365-8711.2001.04493.x}, \href
  {https://ui.adsabs.harvard.edu/abs/2001MNRAS.326..901F} {326, 901}

\bibitem[\protect\citeauthoryear{{Freire}, {Camilo}, {Kramer}, {Lorimer},
  {Lyne}, {Manchester}  \& {D'Amico}}{{Freire} et~al.}{2003}]{fck+03}
{Freire} P.~C.,  {Camilo} F.,  {Kramer} M.,  {Lorimer} D.~R.,  {Lyne} A.~G.,
  {Manchester} R.~N.,   {D'Amico} N.,  2003, \mn@doi [\mnras]
  {10.1046/j.1365-8711.2003.06392.x}, \href
  {https://ui.adsabs.harvard.edu/abs/2003MNRAS.340.1359F} {340, 1359}

\bibitem[\protect\citeauthoryear{{Freire} et~al.,}{{Freire}
  et~al.}{2017}]{frk+17}
{Freire} P.~C.~C.,  et~al., 2017, \mn@doi [\mnras] {10.1093/mnras/stx1533},
  \href {https://ui.adsabs.harvard.edu/abs/2017MNRAS.471..857F} {471, 857}

\bibitem[\protect\citeauthoryear{{Gaia Collaboration} et~al.,}{{Gaia
  Collaboration} et~al.}{2018}]{hvm+18}
{Gaia Collaboration} et~al., 2018, \mn@doi [\aap]
  {10.1051/0004-6361/201832698}, \href
  {https://ui.adsabs.harvard.edu/abs/2018A&A...616A..12G} {616, A12}

\bibitem[\protect\citeauthoryear{{Gnedin}, {Zhao}, {Pringle}, {Fall}, {Livio}
  \& {Meylan}}{{Gnedin} et~al.}{2002}]{gzp+02}
{Gnedin} O.~Y.,  {Zhao} H.,  {Pringle} J.~E.,  {Fall} S.~M.,  {Livio} M.,
  {Meylan} G.,  2002, \mn@doi [\apjl] {10.1086/340319}, \href
  {https://ui.adsabs.harvard.edu/abs/2002ApJ...568L..23G} {568, L23}

\bibitem[\protect\citeauthoryear{{Harris}}{{Harris}}{2010}]{har10}
{Harris} W.~E.,  2010, arXiv e-prints, \href
  {https://ui.adsabs.harvard.edu/abs/2010arXiv1012.3224H} {p. arXiv:1012.3224}

\bibitem[\protect\citeauthoryear{{Harrison} \& {Tademaru}}{{Harrison} \&
  {Tademaru}}{1975}]{Harrison75}
{Harrison} E.~R.,  {Tademaru} E.,  1975, \mn@doi [\apj] {10.1086/153907}, \href
  {https://ui.adsabs.harvard.edu/abs/1975ApJ...201..447H} {201, 447}

\bibitem[\protect\citeauthoryear{{Henleywillis}, {Cool}, {Haggard}, {Heinke},
  {Callanan}  \& {Zhao}}{{Henleywillis} et~al.}{2018}]{hch+18}
{Henleywillis} S.,  {Cool} A.~M.,  {Haggard} D.,  {Heinke} C.,  {Callanan} P.,
   {Zhao} Y.,  2018, \mn@doi [\mnras] {10.1093/mnras/sty675}, \href
  {https://ui.adsabs.harvard.edu/abs/2018MNRAS.479.2834H} {479, 2834}

\bibitem[\protect\citeauthoryear{{Hobbs}, {Edwards}  \& {Manchester}}{{Hobbs}
  et~al.}{2006}]{hem06}
{Hobbs} G.~B.,  {Edwards} R.~T.,   {Manchester} R.~N.,  2006, \mn@doi [\mnras]
  {10.1111/j.1365-2966.2006.10302.x}, \href
  {https://ui.adsabs.harvard.edu/abs/2006MNRAS.369..655H} {369, 655}

\bibitem[\protect\citeauthoryear{{Hobbs} et~al.,}{{Hobbs}
  et~al.}{2020}]{hmd+20}
{Hobbs} G.,  et~al., 2020, \mn@doi [\pasa] {10.1017/pasa.2020.2}, \href
  {https://ui.adsabs.harvard.edu/abs/2020PASA...37...12H} {37, e012}

\bibitem[\protect\citeauthoryear{{Hotan}, {van Straten}  \&
  {Manchester}}{{Hotan} et~al.}{2004}]{hvm04}
{Hotan} A.~W.,  {van Straten} W.,   {Manchester} R.~N.,  2004, \mn@doi [\pasa]
  {10.1071/AS04022}, \href
  {https://ui.adsabs.harvard.edu/abs/2004PASA...21..302H} {21, 302}

\bibitem[\protect\citeauthoryear{{Hui}, {Cheng}  \& {Taam}}{{Hui}
  et~al.}{2010}]{hct10}
{Hui} C.~Y.,  {Cheng} K.~S.,   {Taam} R.~E.,  2010, \mn@doi [\apj]
  {10.1088/0004-637X/714/2/1149}, \href
  {https://ui.adsabs.harvard.edu/abs/2010ApJ...714.1149H} {714, 1149}

\bibitem[\protect\citeauthoryear{{Kalapotharakos}, {Wadiasingh}, {Harding}  \&
  {Kazanas}}{{Kalapotharakos} et~al.}{2021}]{Kalapotharakos21}
{Kalapotharakos} C.,  {Wadiasingh} Z.,  {Harding} A.~K.,   {Kazanas} D.,  2021,
  \mn@doi [\apj] {10.3847/1538-4357/abcec0}, \href
  {https://ui.adsabs.harvard.edu/abs/2021ApJ...907...63K} {907, 63}

\bibitem[\protect\citeauthoryear{{Kerr}}{{Kerr}}{2011}]{KerrWeighted}
{Kerr} M.,  2011, \mn@doi [\apj] {10.1088/0004-637X/732/1/38}, \href
  {http://adsabs.harvard.edu/abs/2011ApJ...732...38K} {732, 38}

\bibitem[\protect\citeauthoryear{{King}}{{King}}{1962}]{kin62}
{King} I.,  1962, \mn@doi [\aj] {10.1086/108756}, \href
  {https://ui.adsabs.harvard.edu/abs/1962AJ.....67..471K} {67, 471}

\bibitem[\protect\citeauthoryear{{Kirsten} et~al.,}{{Kirsten}
  et~al.}{2022}]{Kirsten22}
{Kirsten} F.,  et~al., 2022, \mn@doi [\nat] {10.1038/s41586-021-04354-w}, \href
  {https://ui.adsabs.harvard.edu/abs/2022Natur.602..585K} {602, 585}

\bibitem[\protect\citeauthoryear{{K{\i}z{\i}ltan}, {Baumgardt}  \&
  {Loeb}}{{K{\i}z{\i}ltan} et~al.}{2017}]{kbl17}
{K{\i}z{\i}ltan} B.,  {Baumgardt} H.,   {Loeb} A.,  2017, \mn@doi [\nat]
  {10.1038/nature21361}, \href
  {https://ui.adsabs.harvard.edu/abs/2017Natur.542..203K} {542, 203}

\bibitem[\protect\citeauthoryear{{Kremer} et~al.,}{{Kremer}
  et~al.}{2020}]{kyr+20}
{Kremer} K.,  et~al., 2020, \mn@doi [\apjs] {10.3847/1538-4365/ab7919}, \href
  {https://ui.adsabs.harvard.edu/abs/2020ApJS..247...48K} {247, 48}

\bibitem[\protect\citeauthoryear{{Perera} et~al.,}{{Perera}
  et~al.}{2017}]{psl+17}
{Perera} B.~B.~P.,  et~al., 2017, \mn@doi [\mnras] {10.1093/mnras/stx501},
  \href {https://ui.adsabs.harvard.edu/abs/2017MNRAS.468.2114P} {468, 2114}

\bibitem[\protect\citeauthoryear{{Prager}, {Ransom}, {Freire}, {Hessels},
  {Stairs}, {Arras}  \& {Cadelano}}{{Prager} et~al.}{2017}]{prf+17}
{Prager} B.~J.,  {Ransom} S.~M.,  {Freire} P. C.~C.,  {Hessels} J. W.~T.,
  {Stairs} I.~H.,  {Arras} P.,   {Cadelano} M.,  2017, \mn@doi [\apj]
  {10.3847/1538-4357/aa7ed7}, \href
  {https://ui.adsabs.harvard.edu/abs/2017ApJ...845..148P} {845, 148}

\bibitem[\protect\citeauthoryear{{Ransom}}{{Ransom}}{2001}]{ran01}
{Ransom} S.~M.,  2001, PhD thesis, Harvard University

\bibitem[\protect\citeauthoryear{{Ransom}, {Eikenberry}  \&
  {Middleditch}}{{Ransom} et~al.}{2002}]{rem02}
{Ransom} S.~M.,  {Eikenberry} S.~S.,   {Middleditch} J.,  2002, \mn@doi [\aj]
  {10.1086/342285}, \href
  {https://ui.adsabs.harvard.edu/abs/2002AJ....124.1788R} {124, 1788}

\bibitem[\protect\citeauthoryear{{Ray} et~al.,}{{Ray} et~al.}{2011}]{Ray2011}
{Ray} P.~S.,  et~al., 2011, \mn@doi [\apjs] {10.1088/0067-0049/194/2/17}, \href
  {http://adsabs.harvard.edu/abs/2011ApJS..194...17R} {194, 17}

\bibitem[\protect\citeauthoryear{{Reynoso-Cordova}, {Burgue{\~n}o},
  {Geringer-Sameth}, {Gonz{\'a}lez-Morales}, {Profumo}  \&
  {Valenzuela}}{{Reynoso-Cordova} et~al.}{2021}]{rbg+21}
{Reynoso-Cordova} J.,  {Burgue{\~n}o} O.,  {Geringer-Sameth} A.,
  {Gonz{\'a}lez-Morales} A.~X.,  {Profumo} S.,   {Valenzuela} O.,  2021,
  \mn@doi [\jcap] {10.1088/1475-7516/2021/02/010}, \href
  {https://ui.adsabs.harvard.edu/abs/2021JCAP...02..010R} {2021, 010}

\bibitem[\protect\citeauthoryear{{Ridolfi} et~al.,}{{Ridolfi}
  et~al.}{2016}]{rft+16}
{Ridolfi} A.,  et~al., 2016, \mn@doi [\mnras] {10.1093/mnras/stw1850}, \href
  {https://ui.adsabs.harvard.edu/abs/2016MNRAS.462.2918R} {462, 2918}

\bibitem[\protect\citeauthoryear{{Shklovskii}}{{Shklovskii}}{1970}]{shk70}
{Shklovskii} I.~S.,  1970, \sovast, \href
  {https://ui.adsabs.harvard.edu/abs/1970SvA....13..562S} {13, 562}

\bibitem[\protect\citeauthoryear{{Soltis}, {Casertano}  \& {Riess}}{{Soltis}
  et~al.}{2021}]{scr21}
{Soltis} J.,  {Casertano} S.,   {Riess} A.~G.,  2021, \mn@doi [\apjl]
  {10.3847/2041-8213/abdbad}, \href
  {https://ui.adsabs.harvard.edu/abs/2021ApJ...908L...5S} {908, L5}

\bibitem[\protect\citeauthoryear{{Song}, {Macias}, {Horiuchi}, {Crocker}  \&
  {Nataf}}{{Song} et~al.}{2021}]{smh+21}
{Song} D.,  {Macias} O.,  {Horiuchi} S.,  {Crocker} R.~M.,   {Nataf} D.~M.,
  2021, \mn@doi [\mnras] {10.1093/mnras/stab2406}, \href
  {https://ui.adsabs.harvard.edu/abs/2021MNRAS.507.5161S} {507, 5161}

\bibitem[\protect\citeauthoryear{{Staveley-Smith}, {Bond}, {Bekki}  \&
  {Westmeier}}{{Staveley-Smith} et~al.}{2022}]{sbb+22}
{Staveley-Smith} L.,  {Bond} E.,  {Bekki} K.,   {Westmeier} T.,  2022, \mn@doi
  [\pasa] {10.1017/pasa.2022.23}, \href
  {https://ui.adsabs.harvard.edu/abs/2022PASA...39...26S} {39, e026}

\bibitem[\protect\citeauthoryear{{Verbunt} \& {Freire}}{{Verbunt} \&
  {Freire}}{2014}]{vf14}
{Verbunt} F.,  {Freire} P. C.~C.,  2014, \mn@doi [\aap]
  {10.1051/0004-6361/201321177}, \href
  {https://ui.adsabs.harvard.edu/abs/2014A&A...561A..11V} {561, A11}

\bibitem[\protect\citeauthoryear{{Wang}, {Bi}  \& {Yin}}{{Wang}
  et~al.}{2021}]{wby21}
{Wang} J.-W.,  {Bi} X.-J.,   {Yin} P.-F.,  2021, \mn@doi [\prd]
  {10.1103/PhysRevD.104.103015}, \href
  {https://ui.adsabs.harvard.edu/abs/2021PhRvD.104j3015W} {104, 103015}

\bibitem[\protect\citeauthoryear{{Ye}, {Kremer}, {Chatterjee}, {Rodriguez}  \&
  {Rasio}}{{Ye} et~al.}{2019}]{ykc+19}
{Ye} C.~S.,  {Kremer} K.,  {Chatterjee} S.,  {Rodriguez} C.~L.,   {Rasio}
  F.~A.,  2019, \mn@doi [\apj] {10.3847/1538-4357/ab1b21}, \href
  {https://ui.adsabs.harvard.edu/abs/2019ApJ...877..122Y} {877, 122}

\bibitem[\protect\citeauthoryear{{Ye}, {Kremer}, {Rodriguez}, {Rui},
  {Weatherford}, {Chatterjee}, {Fragione}  \& {Rasio}}{{Ye}
  et~al.}{2022}]{ykr+22}
{Ye} C.~S.,  {Kremer} K.,  {Rodriguez} C.~L.,  {Rui} N.~Z.,  {Weatherford}
  N.~C.,  {Chatterjee} S.,  {Fragione} G.,   {Rasio} F.~A.,  2022, \mn@doi
  [\apj] {10.3847/1538-4357/ac5b0b}, \href
  {https://ui.adsabs.harvard.edu/abs/2022ApJ...931...84Y} {931, 84}

\bibitem[\protect\citeauthoryear{{Zhao} \& {Heinke}}{{Zhao} \&
  {Heinke}}{2022}]{zh22}
{Zhao} J.,  {Heinke} C.~O.,  2022, \mn@doi [\mnras] {10.1093/mnras/stac442},
  \href {https://ui.adsabs.harvard.edu/abs/2022MNRAS.511.5964Z} {511, 5964}

\bibitem[\protect\citeauthoryear{{Zocchi}, {Gieles}  \&
  {H{\'e}nault-Brunet}}{{Zocchi} et~al.}{2019}]{zgh19}
{Zocchi} A.,  {Gieles} M.,   {H{\'e}nault-Brunet} V.,  2019, \mn@doi [\mnras]
  {10.1093/mnras/sty1508}, \href
  {https://ui.adsabs.harvard.edu/abs/2019MNRAS.482.4713Z} {482, 4713}

\bibitem[\protect\citeauthoryear{{de Jager}, {Raubenheimer}  \&
  {Swanepoel}}{{de Jager} et~al.}{1989}]{Htest}
{de Jager} O.~C.,  {Raubenheimer} B.~C.,   {Swanepoel} J.~W.~H.,  1989, \aap,
  \href {https://ui.adsabs.harvard.edu/abs/1989A&A...221..180D} {221, 180}

\bibitem[\protect\citeauthoryear{{van Straten} \& {Bailes}}{{van Straten} \&
  {Bailes}}{2011}]{vb11}
{van Straten} W.,  {Bailes} M.,  2011, \mn@doi [\pasa] {10.1071/AS10021}, \href
  {https://ui.adsabs.harvard.edu/abs/2011PASA...28....1V} {28, 1}

\makeatother
\end{thebibliography}




\bsp	
\label{lastpage}
\end{document}